\documentclass[aps,twocolumn,pra,superscriptaddress,showpacs,tightenlines]{revtex4-1}
\usepackage{amsmath}
\usepackage{graphicx}
\usepackage{color}
\usepackage{amsfonts}
\usepackage{txfonts}
\usepackage[colorlinks,citecolor=blue]{hyperref}
\begin{document}

\title{Generation of macroscopic Schr\"{o}dinger-cat states in qubit-oscillator systems}
\author{Jie-Qiao Liao}
\email{jieqiaoliao@gmail.com}
\affiliation{School of Natural
Sciences, University of California, Merced, California 95343, USA}
\author{Jin-Feng Huang}
\affiliation{Key Laboratory of Low-Dimensional Quantum Structures and Quantum Control of Ministry of Education, Department of Physics and Synergetic Innovation Center for Quantum Effects
and Applications, Hunan Normal University, Changsha 410081, China}
\author{Lin Tian}
\email{ltian@ucmerced.edu}
\affiliation{School of Natural
Sciences, University of California, Merced, California 95343, USA}

\begin{abstract}
We propose a scheme to generate macroscopic Schr\"{o}dinger-cat states in a quantum harmonic oscillator (electromagnetic field or mechanical resonator) coupled to a quantum bit (two-level system) via a conditional displacement mechanism. By driving the qubit monochromatically, the oscillation of the qubit state modifies the effective frequency of the driving force acting on the oscillator, and a resonant or near-resonant driving on the oscillator can be achieved. The displacement of the oscillator is then significantly enhanced due to the small detuning of the driving force and can exceed that of the zero-point fluctuation. This effect can be used to prepare quantum superpositions of macroscopically distinct coherent states in the oscillator. We present detailed studies on this state-generation scheme in both the closed- and open-system cases. This approach can be implemented in various experimental platforms, such as cavity- or circuit-QED systems, electromechanical systems, and spin-cantilever systems.
\end{abstract}

%42.50.Dv     Quantum state engineering and measurements
%42.50.Pq     Cavity quantum electrodynamics; micromasers
%03.65.Yz     Decoherence; open systems; quantum statistical methods

\maketitle

\section{Introduction\label{sec:intro}}

The quantum superposition principle~\cite{Dirac1958}, one of the cornerstones of quantum theory, has attracted heavy attention from both theorists and experimentalists since its discovery. Researchers have put intensive effort into studying the generation of quantum superposition states. Such states play an important role in the study of the foundations of quantum theory~\cite{Leggett2002}, giving insight into the questions of quantum-classical boundary and quantum decoherence~\cite{Zurek1991}. Quantum superposition has been observed in numerous physical systems~\cite{Arndt2014,Nimmrichter2013}, including electronic~\cite{Clarke1988,Friedman2000,Mooij2000}, photonic~\cite{Brune1996,Auffeves2003,Polzik2006,Ourjoumtsev2006,Ourjoumtsev2006b,Wakui2007,Haroche2008,Schoelkopf2013,Devoret2015}, and atomic degrees of freedom~\cite{Monroe1996,Zeilinger1999}. Nevertheless, it remains a big challenge to create superpositions of macroscopically distinct coherent states in nanomechanical systems~\cite{Huang2003,Ekinci2005,Schwab2005,Cleland2010,Blencowe2004,Poot2012,Xiang2013}, in part due to the difficulty in generating coherent states with macroscopically distinct amplitudes in the phase space in such systems.

The conditional displacement mechanism is an important method for generating the Schr\"{o}dinger cat states~\cite{Monroe1996}. In this method, the magnitude of the coherent states is determined by the ratio of the conditional coupling strength over the frequency of the oscillator. Consider a coupled qubit-oscillator system described by the Hamiltonian ${\cal H}=\omega_{r}a^{\dagger}a+g_{0}\sigma_{z}(a+a^{\dagger})$ [cf. Eq.~(\ref{Hamilt1}) for the notations]. Corresponding to the two eigenstates of $\sigma_{z}$, the interaction $g_{0}\sigma_{z}(a+a^{\dagger})$ between the qubit and the oscillator produces displacements with the same maximum magnitude $2g_{0}/\omega_{r}$ but opposite direction in the phase space of the oscillator. The conditional dynamics in this system can then be used to create superposed coherent states in the oscillator~\cite{Armour2002,Marshall2003}. In current experiments, the coupling strength $g_{0}$ is much smaller than the frequency $\omega_{r}$ of the oscillator. Even in qubit-oscillator systems with ultrastrong coupling~\cite{Ashhab2010,Niemczyk2010}, $g_{0}/\omega_{r}$ (its value is usually $\geq0.1$) is still smaller than $1$. Therefore, the displacement of the oscillator will be shadowed by its zero-point fluctuation, and one cannot see clear evidence of quantum superposition.

To date, many approaches have been proposed to create superposed coherent states in various setups~\cite{Milburn1986,Yurke1986,Miranowicz1990,Gea-Banacloche1991,Brune1992,Buzek1992a,Davidovich1996,Zheng1998,Gerry1999,Liu2005,Liao2008,Cirac2011,Yin2013,Tan2013,Ge2015,Zhang2015}. In particular, several approaches have been proposed to enhance the displacement of mechanical resonators so that macroscopically distinct superposed coherent states can be observed. For example, one of us (Tian)~\cite{Tian2005} proposed a scheme to increase the mechanical displacement in a nanomechanical system by introducing a series of $\pi$ pulses to flip the state of a superconducting qubit. By flipping the qubit state at properly selected times, the total mechanical displacement can be accumulated and be significantly amplified. In \cite{Kolkowitz2012,Bennett2012}, the authors have proposed to enhance the mechanical displacement of a cantilever in a spin-cantilever system by applying an XY4 pulse sequence to the spin. In addition, in \cite{Liao2015}, one of us (Liao) proposed a scheme to enhance the mechanical displacement of a single photon by introducing cavity frequency modulation in an optomechanical system with the ``membrane-in-the-middle" configuration.

In this paper, we propose an efficient approach for creating quantum superpositions of large-amplitude coherent states in a qubit-oscillator system by applying a monochromatic driving to the qubit. Under appropriate conditions, this system can be described by an approximate Hamiltonian that depicts a conditionally driven quantum harmonic oscillator. In this Hamiltonian, the effective detuning can be tuned to be much smaller than the magnitude of the driving. Consequently, the displacement of the oscillator can be enhanced significantly to be larger than the zero-point motion. It is thus promising to observe macroscopically distinct superpositions of coherent states in such systems. One advantage of this method is that it does not require accurate control of the exact wave form of the driving pulses, as only a sinusoidal driving on the qubit is used.

The rest of this paper is organized as follows. In Sec.~\ref{sec:model}, we introduce the qubit-oscillator system and derive an approximate Hamiltonian of this system under driving. In Sec.~\ref{sec:stategeneration}, we study the generation of macroscopically distinct Schr\"{o}dinger-cat states with this approximate Hamiltonian and verify the validity of the rotating-wave approximation. We also study the quantum entanglement between the qubit and the oscillator. Moreover, we investigate the Wigner function and the probability distribution of the rotated quadrature operator to study the quantum interference and coherence in the generated states. In Sec.~\ref{sec:opensys}, we discuss the open-system case for this state-generation scheme. We study the influence of the dissipations on the fidelity, the probability, the Wigner function, and the probability distribution of the rotated quadrature operator. In Sec.~\ref{sec:othercouplings}, we show that the current method can be extended to various other forms of qubit-oscillator coupling. Finally, we present discussions in Sec.~\ref{sec:discussion} and conclusions in Sec.~\ref{sec:conclusion}.

\section{System and Hamiltonian \label{sec:model}}
%%%%%%%%%%%%%%%%%%%%%%%
\begin{figure}[tbp]
\center
\includegraphics[bb=32 635 367 772, width=0.47 \textwidth]{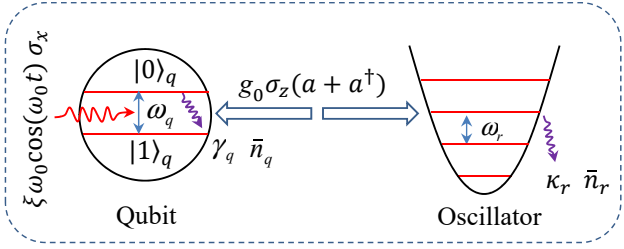}
\caption{Schematic of a coupled qubit-oscillator system. The qubit with an energy separation $\omega_{q}$ is coupled to a quantum harmonic oscillator (represented by a harmonic trap) with a resonance frequency $\omega_{r}$ via a conditional displacement interaction $g_{0}\sigma_{z}(a+a^{\dagger})$. Along the $x$-direction in the Bloch sphere, the qubit is driven by a monochromatic field with a frequency $\omega_{0}$ and a magnitude $\xi\omega_{0}$. The decay rate and the thermal excitation occupation number of the qubit (oscillator) are denoted by $\gamma_{q}$ ($\kappa_{r}$) and $\bar{n}_{q}$ ($\bar{n}_{r}$), respectively.}
\label{Fig1}
\end{figure}
%%%%%%%%%%%%%%%%%%%%%%%

The coupled qubit-oscillator system (Fig.~\ref{Fig1}) is described by the Hamiltonian ($\hbar=1$)
\begin{equation}
H(t)=\frac{\omega_{q}}{2}\sigma_{z}+\xi\omega_{0}\cos(\omega_{0}t)\sigma_{x}+\omega_{r}a^{\dagger}a+g_{0}\sigma_{z}\left(a+a^{\dagger}\right),\label{Hamilt1}
\end{equation}
where $a$ $(a^{\dagger})$ is the annihilation (creation) operator of the quantum harmonic oscillator (electromagnetic field or mechanical resonator) with frequency $\omega_{r}$. The quantum bit (two-level system) is described by the Pauli operators $\sigma_{x}=\vert 0\rangle_{q}\!\,_{q}\langle 1\vert +\vert 1\rangle_{q}\!\,_{q}\langle 0\vert$, $\sigma_{y}= i(\vert 1\rangle_{q}\!\,_{q}\langle 0\vert-\vert 0\rangle_{q}\!\,_{q}\langle 1\vert)$, and $\sigma_{z}=\vert 0\rangle_{q}\!\,_{q}\langle 0\vert -\vert 1\rangle_{q}\!\,_{q}\langle 1\vert$, where $\vert 0\rangle_{q}$ and $\vert 1\rangle_{q}$ are, respectively, the excited state and the ground state, with an energy separation $\omega_{q}$. The qubit is driven by a monochromatic field with a dimensionless driving amplitude $\xi$ and a frequency $\omega_{0}$. The $g_{0}$ term  is the conditional displacement interaction between the qubit and the oscillator. This Hamiltonian can be realized in various experimental platforms, such as cavity- or circuit-QED systems~\cite{Wallraff2004}, electromechanical systems~\cite{Cleland2010,LaHaye2009}, and spin-oscillator systems~\cite{Kolkowitz2012, Rugar2004}.

To study the impact of the qubit driving on the dynamics of the system, we perform the transformation
\begin{equation}
V(t)=\exp\left\{-i\left[\xi\sin(\omega_{0}t)\sigma_{x}+\omega_{r}ta^{\dagger}a\right]\right\}\label{transfV}
\end{equation}
on this coupled system. In the rotating frame defined by $V(t)$, the transformed Hamiltonian becomes
\begin{eqnarray}
\tilde{H}(t)&=&V^{\dag}(t)H(t)V(t)-iV^{\dag}(t)\dot{V}(t)\nonumber\\
&=&\frac{\omega_{q}}{2}\left[\cos[2\xi\sin(\omega_{0}t)]\sigma_{z}+\sin[2\xi\sin(\omega_{0}t)]\sigma_{y}\right]\nonumber\\
&&+g_{0}\left[\cos[2\xi\sin(\omega_{0}t)]\sigma_{z}+\sin[2\xi\sin(\omega_{0}t)]\sigma_{y}\right]\nonumber\\
&&\times\left(ae^{-i\omega_{r}t}+a^{\dagger}e^{i\omega_{r}t}\right).
\end{eqnarray}
Under the Jacobi-Anger expansions
\begin{subequations}
\label{JAexpansion}
\begin{align}
\cos[2\xi\sin(\omega_{0}t)]=&J_{0}(2\xi)+2\sum_{n=1}^{\infty}J_{2n}(2\xi)\cos(2n\omega_{0}t),\\
\sin[2\xi\sin(\omega_{0}t)]=&2\sum_{n=1}^{\infty}J_{2n-1}(2\xi)\sin[(2n-1)\omega_{0}t],
\end{align}
\end{subequations}
with $J_{m}(x)$ being the Bessel function of the first kind and $m$ being an integer, the Hamiltonian $\tilde{H}(t)$ can be expanded into
\begin{eqnarray}
\tilde{H}(t)&=&\frac{\omega_{q}}{2}J_{0}(2\xi)\sigma_{z}
+\sum_{n=1}^{\infty}\omega_{q}\left[J_{2n}(2\xi)\cos(2n\omega_{0}t)\sigma_{z}\right.\nonumber\\
&&\left.+J_{2n-1}(2\xi)\sin[(2n-1)\omega_{0}t]\sigma_{y}\right]\nonumber\\
&&+g_{0}J_{0}(2\xi)\sigma_{z}\left(ae^{-i\omega_{r}t}+a^{\dagger}e^{i\omega_{r}t}\right)\nonumber\\
&&+\sum_{n=1}^{\infty}g_{0}\left[\underline{J_{2n}(2\xi)\sigma_{z}\left(ae^{-i(\omega_{r}-2n\omega_{0})t}+\textrm{H.c.}\right)}\right.\nonumber\\
&&\left.+J_{2n}(2\xi)\sigma_{z}\left(ae^{-i(\omega_{r}+2n\omega_{0})t}+\textrm{H.c.}\right)\right.\nonumber\\
&&\left.-iJ_{2n-1}(2\xi)\sigma_{y}\left(ae^{-i[\omega_{r}-(2n-1)\omega_{0}]t}-\textrm{H.c.}\right)\right.\nonumber\\
&&\left.+iJ_{2n-1}(2\xi)\sigma_{y}\left(ae^{-i[\omega_{r}+(2n-1)\omega_{0}]t}-\textrm{H.c.}\right)\right].
\label{Hsummation}
\end{eqnarray}
This Hamiltonian contains oscillating terms that differ by frequencies of $m\omega_{0}$, with $m$ being an integer. We consider the case $\omega_{r}\gg g_{0}>g_{0}J_{0}(2\xi)$ and choose a driving frequency $\omega_{0}$ that satisfies $\omega_{0}\gg g_{0}>g_{0}J_{2n}(2\xi)$, $g_{0}J_{2n-1}(2\xi)$. Furthermore, it also ensures that there is a near-resonant term (corresponding to a characteristic number $n_{0}$) in the form of $g\sigma_{z}(ae^{-i\delta t}+a^{\dagger }e^{i\delta t})$ [cf. Eq.~(\ref{Hamappr})] in the fourth line of Eq.~(\ref{Hsummation}) (underlined). Here, $\delta$ is an effective driving detuning and $g$ is the normalized coupling coefficient with
\begin{equation}
\delta=\omega_{r}-2n_{0}\omega_{0}, \hspace{1 cm} g=g_{0}J_{2n_{0}}(2\xi).\label{deltaandg}
\end{equation}
For a given $\omega_{0}$, the parameter $n_{0}$ should be chosen such that the corresponding $J_{2n_{0}}(2\xi)$ term is the nearest-resonant term, i.e., $n_{0}=\texttt{Round}[\omega_{r}/(2\omega_{0})]$, where $\texttt{Round}[x]$ is a function for getting the nearest integer of $x$.

When the characteristic number $n_{0}$ is chosen, we can tune $\omega_{0}$ such that $\delta$ can be comparable or even smaller than the coupling coefficient $g$, whereas all other $g_{0}$ terms are fast-oscillating terms that can be omitted under the rotating-wave approximation (RWA). Moreover, by assuming $\omega_{0}\gg \omega_{q}J_{2n}(2\xi)/2,\omega_{q}J_{2n-1}(2\xi)/2$, only the term $\omega_{q}J_{0}(2\xi)\sigma_{z}/2$ in all $\omega_{q}$ terms needs to be preserved. This term will introduce an additional phase factor to the state of the qubit, but it will not affect the dynamics of the oscillator. Hereafter, we assume $\omega_{q}=0$ for simplicity of discussion. Hence under the condition
\begin{equation}
\vert\delta\vert,\; g_{0}\ll \omega_{0},\; \omega_{r},\label{RWAcond}
\end{equation}
the high-frequency oscillating terms in Eq.~(\ref{Hsummation}) can be neglected by applying the RWA, and we obtain the approximate Hamiltonian
\begin{equation}
\tilde{H}_{\mbox{\scriptsize RWA}}(t)=g\sigma_{z}\left(ae^{-i\delta t}+a^{\dagger}e^{i\delta t}\right).\label{Hamappr}
\end{equation}
This Hamiltonian describes a quantum harmonic oscillator that is conditionally displaced by the states of the qubit. When the qubit is prepared in the eigenstates $|0\rangle_{q}$ and $|1\rangle_{q}$ of the $\sigma_{z}$ operator ($\sigma_{z}|0\rangle_{q}=|0\rangle_{q}$ and $\sigma_{z}|1\rangle_{q}=-|1\rangle_{q}$), the displacement forces acting on the oscillator are in the opposite directions. Therefore, corresponding to the qubit's states $|0\rangle_{q}$ and $|1\rangle_{q}$, if the oscillator is initially prepared in its ground state, its states at time $t$ would be coherent states with the same magnitude but opposite phase in the phase space [as shown in the second line of Eq.~(\ref{statephiss})]. From Eq.~(\ref{deltaandg}), we see that the magnitude of $g$ can be maximized by optimizing the value of $\xi$ so that $J_{2n_{0}}(2\xi)$ is at its peak values. In the following simulations, we choose $n_{0}=1$ and $\xi=1.5271$, which corresponds to the first peak value of $J_{2}(2\xi)$. We choose a small $\delta$ by adjusting $\omega_{0}$, which strongly enhances the displacement of the oscillator. For the generation of macroscopically distinct coherent-state components~\cite{Yurke1986}, i.e., $|\alpha|_{\textrm{max}}>1$ [cf. Eq.~(\ref{expofalpha})], the detuning should satisfy the condition $\delta<2g$ (hereafter, we assume $\delta>0$). By preparing the qubit in a superposition of its two eigenstates, this conditional dynamics can be used to create macroscopic superpositions of large-amplitude coherent states in the oscillator. The generated macroscopic cat states are useful for probing macroscopic realism in massive systems~\cite{Asadian2014}.

\section{Generation of Schrodinger's cat states\label{sec:stategeneration}}

In this section, we study the dynamics of the coupled system described in Sec.~\ref{sec:model}. We also discuss the validity of the RWA condition (\ref{RWAcond}) by examining the fidelity between an approximate state and the exact state, which
evolve under the approximate Hamiltonian~(\ref{Hamappr}) and the full Hamiltonian~(\ref{Hamilt1}), respectively. Moreover, the quantum interference and coherence effects in the generated cat states will be studied.

\subsection{Analytical solution under the RWA\label{ssec:analytical}}

Denoting the state of the system in the original (Schr\"{o}dinger) representation as $\vert\psi(t)\rangle$ and the state in the rotating representation defined by $V(t)$ as $\vert\phi(t)\rangle$, we have the relations $\vert\psi(t)\rangle=V(t)\vert\phi(t)\rangle$ and $\vert\psi(0)\rangle=\vert\phi(0)\rangle$. Using the Magnus expansion, the unitary evolution operator associated with the Hamiltonian $\tilde{H}_{\mbox{\scriptsize RWA}}(t)$ can be expressed as (see the Appendix)
\begin{equation}
U(t)=\exp\left[i\theta(t)\right]\exp\left\{\sigma_{z}\left[\eta(t)a^{\dagger}-\eta^{\ast}(t)a\right]\right\},\label{unitaryoptana}
\end{equation}
where $\theta(t)=(g/\delta)^2 [\delta t-\sin(\delta t)]$ is a global phase factor and $\eta(t)=(g/\delta)(1-e^{i\delta t})$ is the displacement amplitude of the oscillator. We consider an initial state $\vert\phi(0)\rangle=\vert+\rangle_{q}\vert 0\rangle_{r}$ of the system, where $\vert+\rangle_{q}=(\vert0\rangle_{q}+\vert 1\rangle_{q})/\sqrt{2}$ is the eigenstate of $\sigma_{x}$ with eigenvalue $+1$, and $\vert 0\rangle_{r}$ is the ground state of the harmonic oscillator. By utilizing the unitary evolution operator $U(t)$, the state of the system at time $t$ can be obtained as
\begin{eqnarray}
\vert\phi(t)\rangle &=&U(t)\vert\phi(0)\rangle\nonumber \\
&=&\frac{e^{i\theta(t)}}{\sqrt{2}}\left(\vert0\rangle_{q}\vert\eta(t)\rangle_{r}+\vert1\rangle_{q}\vert-\eta(t)\rangle_{r}\right)\nonumber \\
&=&\frac{e^{i\theta(t)}}{2}\left[\vert+\rangle_{q}\left(\vert\eta(t)\rangle_{r}+\vert-\eta(t)\rangle_{r}\right)\right.\nonumber\\
&&\left.+\vert-\rangle_{q}\left(\vert\eta(t)\rangle_{r}-\vert-\eta(t)\rangle_{r}\right)\right],\label{statephiss}
\end{eqnarray}
where $|\eta(t)\rangle_{r}$ and $|-\eta(t)\rangle_{r}$ are coherent states of the harmonic oscillator, with the same amplitude but opposite phase in the phase space. Using the transformation $V(t)$, the corresponding state in the original representation can be expressed as
\begin{eqnarray}
\vert\psi(t)\rangle &=&\frac{e^{i\theta(t)}}{2}\left[
\mathcal{N}^{-1}_{+}(t)e^{-i\xi\sin(\omega_{0}t)}\vert +\rangle_{q}\vert \alpha_{+}(t)\rangle_{r}\right.\nonumber\\
&&\left.+\mathcal{N}^{-1}_{-}(t)e^{i\xi\sin(\omega_{0}t)}\vert-\rangle_{q}\vert\alpha_{-}(t)\rangle_{r}\right],
\label{statepsi}
\end{eqnarray}
where we introduced the even and odd coherent states (the Schr\"{o}dinger cat states)~\cite{Dodonov1974},
\begin{equation}
\vert\alpha_{\pm}(t)\rangle_{r}=\mathcal{N}_{\pm}(t)(\vert\alpha(t)\rangle_{r}\pm\vert-\alpha(t)\rangle_{r}),\label{catstates}
\end{equation}
with normalization constants
\begin{equation}
\mathcal{N}_{\pm}(t)=\left[2\left(1\pm e^{-2\vert\alpha(t)\vert^{2}}\right)\right]^{-1/2},\label{normconstNpm}
\end{equation}
and coherent-state amplitude
\begin{equation}
\alpha(t)=\frac{g}{\delta}\left(1-e^{i\delta t}\right)e^{-i\omega_{r}t}.\label{expofalpha}
\end{equation}
For the state $\vert\psi(t)\rangle$, the average excitation number in the oscillator is
\begin{equation}
\langle\psi(t)\vert a^{\dagger}a\vert\psi(t)\rangle=\frac{4 g^{2}}{\delta^{2}}\sin^{2}\left(\frac{\delta t}{2}\right),\label{avphononanal}
\end{equation}
which is the absolute square  $|\alpha(t)|^{2}$ of the coherent-state amplitude. Equation~(\ref{expofalpha}) shows that the maximum displacement amplitude is $\vert\alpha\vert_{\mbox{\scriptsize max}}=2g/\delta$, and it can be obtained at the times $ t=(2m+1)\pi/\delta$ for natural numbers $m$. By choosing a small $\delta$ value, we can create macroscopically distinct Schr\"{o}dinger-cat states with $|\alpha|>1$. When $\delta=0$, the oscillator is driven resonantly and the displacement amplitude becomes $\alpha_{\textrm{res}}(t)=-igt\exp(-i\omega_{r}t)$, which increases linearly in time. The damping of the oscillator and the finite duration of the evolution will prevent the system from diverging into instability.

To generate the Schr\"{o}dinger-cat states~(\ref{catstates}), we perform a qubit measurement on the state $\vert\psi(t)\rangle$. When the $\sigma_{x}$ operator of the qubit is detected (i.e., in the bases of $\vert \pm\rangle_{q}$), the oscillator will collapse into the Schr\"{o}dinger-cat states $\vert\alpha_{\pm}(t)\rangle_{r}$. The probabilities of the states $\vert\alpha_{\pm}(t)\rangle_{r}$ are
\begin{eqnarray}
\mathcal{P}_{\pm}(t)=\frac{1}{2}\left(1\pm e^{-2\vert\alpha(t)\vert^{2}}\right),\label{Probanaly}
\end{eqnarray}
which are determined by $g$ and $\delta$, but independent of $\omega_{r}$.

\subsection{Entanglement between the qubit and the oscillator}

The state $\vert\psi(t)\rangle$ in Eq.~(\ref{statepsi}) is an entangled state of the coupled qubit-oscillator system. For this bipartite pure state, the degree of entanglement can be characterized with the von Neumann entanglement entropy~\cite{Bennett1996},
\begin{equation}
S=-\textrm{Tr}\left[\rho_{q}\log_{2}\rho_{q}\right]=-\textrm{Tr}\left[\rho_{r}\log_{2}\rho_{r}\right],
\end{equation}
where $\rho_{q}$ ($\rho_{r}$) is the reduced density matrix of the qubit (oscillator).
Using Eq.~(\ref{statepsi}), we obtain the density matrix $\rho_{q}$ as
\begin{equation}
\rho_{q}(t)=\frac{1}{4\mathcal{N}^{2}_{+}(t)}\vert +\rangle_{q}\;_{q}\langle +|+\frac{1}{4\mathcal{N}^{2}_{-}(t)}\vert -\rangle_{q}\;_{q}\langle -|.
\end{equation}
The von Neumann entropy is then
\begin{eqnarray}
S(t)=\frac{1}{4\mathcal{N}^{2}_{+}(t)}\log_{2}\left[4\mathcal{N}^{2}_{+}(t)\right]
+\frac{1}{4\mathcal{N}^{2}_{-}(t)}\log_{2}\left[4\mathcal{N}^{2}_{-}(t)\right],
\end{eqnarray}
where $\mathcal{N}_{\pm}(t)$ are given by Eq.~(\ref{normconstNpm}).

%%%%%%%%%%%%%%%%%%%%%%%
\begin{figure}[tbp]
\center
\includegraphics[bb=21 57 340 267, width=0.47 \textwidth]{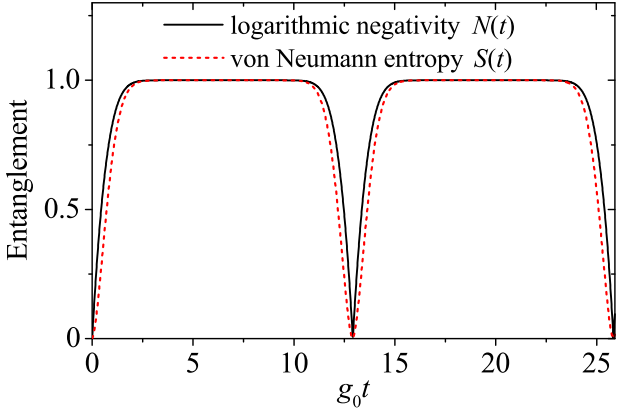}
\caption{Time dependence of the entanglement of the state $\vert\psi(t)\rangle$ in Eq.~(\ref{statepsi}): the von Neumann entropy $S(t)$ (red short-dashed curve) and the logarithmic negativity $N(t)$ (black solid curve). The used parameters are: $n_{0}=1$, $\xi=1.5271$, and $\delta=g$.}
\label{Fig2}
\end{figure}
%%%%%%%%%%%%%%%%%%%%%%%
It should be pointed out that the von Neumann entropy cannot be used to quantify the bipartite entanglement of mixed states.
To consistently describe the bipartite entanglement in both the closed- and open-system cases, we introduce the logarithmic negativity~\cite{Vidal2002,Plenio2005}, which is defined by
\begin{equation}
N=\log_{2}\Vert \rho^{T_{r}}\Vert_{1},\label{lognegtivity}
\end{equation}
where $T_{r}$ denotes the partial transpose of the density matrix $\rho$ of the system with respect to the oscillator, and the trace norm $\Vert \rho^{T_{r}}\Vert_{1}$ is defined by
\begin{equation}
\Vert \rho^{T_{r}}\Vert_{1} =\textrm{Tr}\left[\sqrt{(\rho^{T_{r}})^{\dagger}\rho^{T_{r}}}\right].
\end{equation}
Using Eq.~(\ref{statepsi}), the logarithmic negativity of the density matrix $\rho$ can be obtained as
\begin{equation}
N(t)=\log_{2}\left[\frac{1}{4}\left(\frac{1}{\mathcal{N}_{+}(t)}+\frac{1}{\mathcal{N}_{-}(t)}\right)^{2}\right].
\end{equation}

In Fig.~\ref{Fig2}, we plot the von Neumann entropy $S(t)$ and the logarithmic negativity $N(t)$ as functions of time $t$.
We can see that the entanglement is a periodic function of time $t$ with the same period as the coherent amplitude $|\alpha(t)|$. At times $t=2m\pi/\delta$ for natural numbers $m$, the coherent amplitude $\alpha(t)=0$. As a result, the qubit and the oscillator decouple and the entanglement disappears. In the middle duration of a period, the von Neumann entropy $S(t)$ and the logarithmic negativity $N(t)$ reach the maximum. In these durations, the coherent amplitude $|\alpha(t)|$ is large enough such that $\exp[-2|\alpha(t)|^{2}]\approx0$ [i.e., $\mathcal{N}_{\pm}(t)\approx1/\sqrt{2}$], and then the state $\vert\psi(t)\rangle$ can be approximated by a Bell-like state defined with the orthogonal basis states $|\pm\rangle_{q}$ and $|\alpha_{\pm}(t)\rangle_{r}$. In the middle duration of a period, the two measures agree well.

\subsection{Numerical solution with the full Hamiltonian\label{ssec:numerical}}

We now calculate the state of this system by solving the full Hamiltonian (\ref{Hamilt1}) numerically. In a closed system without decoherence from the qubit and the harmonic oscillator, a pure state of the system has the general form
\begin{equation}
\vert\Psi(t)\rangle=\sum_{m=0}^{\infty}\left[A_{m}(t)\vert 0\rangle_{q}\vert m\rangle_{r}+B_{m}(t)\vert1\rangle_{q}\vert m\rangle_{r}\right],\label{statePsi}
\end{equation}
where $\vert m\rangle_{r}$ are the Fock states of the oscillator. Following the Schr\"{o}dinger equation under the Hamiltonian $H(t)$, the equations of motion for the probability amplitudes $A_{m}(t)$ and $B_{m}(t)$ (for natural numbers $m$) can be derived as
\begin{subequations}
\begin{align}
\dot{A}_{m}(t)=&-i\xi\omega_{0}\cos(\omega_{0}t)B_{m}(t)-im\omega_{r}A_{m}(t)\nonumber\\
&-ig_{0}\left[\sqrt{m+1}A_{m+1}(t)+\sqrt{m}A_{m-1}(t)\right],\label{eqofmotproampa}\\
\dot{B}_{m}(t)=&-i\xi \omega_{0}\cos(\omega_{0}t)A_{m}(t)-im\omega_{r}B_{m}(t)\nonumber\\
&+ig_{0}\left[\sqrt{m+1}B_{m+1}(t)+\sqrt{m}B_{m-1}(t)\right].\label{eqofmotproampb}
\end{align}
\end{subequations}

For the initial state $\vert+\rangle_{q}\vert 0\rangle_{r}$, we have $A_{0}(0)=B_{0}(0)=1/\sqrt{2}$,  $A_{m>0}(0)=0$, and $B_{m>0}(0)=0$. By numerically solving Eqs.~(\ref{eqofmotproampa}) and (\ref{eqofmotproampb}) with these initial conditions, the evolution of the probability amplitudes can be obtained. In realistic simulations, we need to truncate the Hilbert space of the oscillator such that the equations of motion~(\ref{eqofmotproampa}) and~(\ref{eqofmotproampb}) are closed. The truncation dimension $n_{d}$ should be chosen to ensure the normalization of the state~(\ref{statePsi}). In our simulations, we consider the case of $\delta=g$, which leads to the maximum coherent amplitude $|\alpha|_{\textrm{max}}=2$. Then we choose the truncation dimension as $n_{d}=14$ in the closed-system case. This value should be increased in the open-system case because the oscillator will be excited by the heat baths.

After a measurement of the qubit $\sigma_{x}$ operator [with eigenstates $|\pm\rangle_{q}=(|0\rangle_{q}\pm|1\rangle_{q})/\sqrt{2}$], the oscillator can be prepared in states
\begin{equation}
\vert\Psi_{\pm}(t)\rangle_{r}=\frac{1}{\sqrt{2p_{\pm}(t)}}\sum_{m=0}^{\infty }\left[A_{m}(t)\pm B_{m}(t)\right]\vert m\rangle_{r},\label{stateexactclo}
\end{equation}
where
\begin{equation}
p_{\pm}(t)=\frac{1}{2}\sum_{m=0}^{\infty}\left\vert A_{m}(t)\pm B_{m}(t)\right\vert^{2}\label{probpmclo}
\end{equation}
are the probabilities of the states $\vert\Psi_{\pm}(t)\rangle_{r}$, respectively.
%%%%%%%%%%%%%%%%%%%%%%%
\begin{figure}[tbp]
\center
\includegraphics[bb= 1 1 275 425, width=0.47 \textwidth]{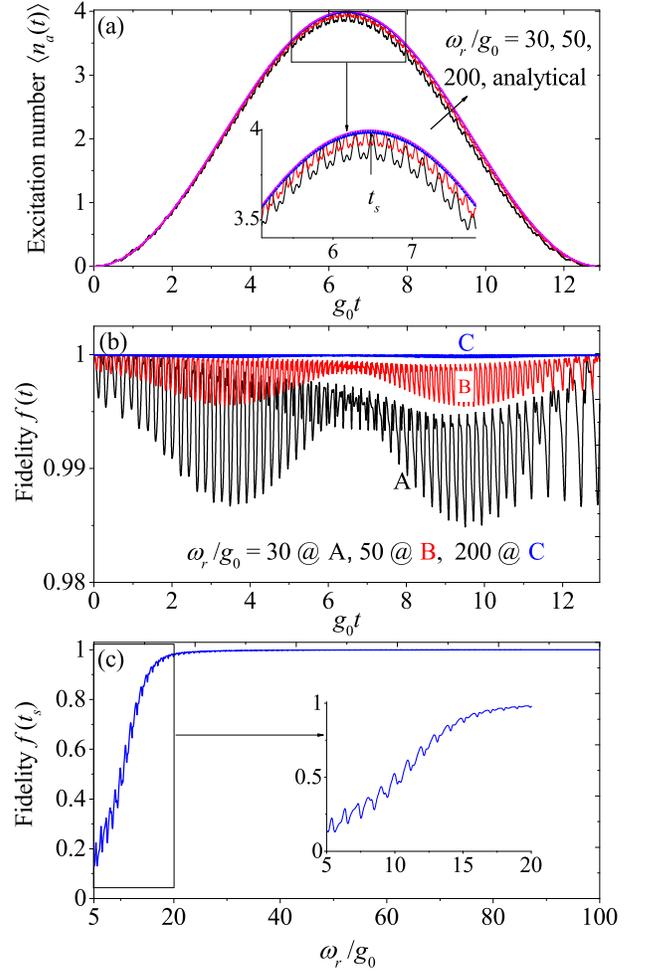}
\caption{(a) Time dependence of the average excitation number $\langle n_{a}(t)\rangle$ in the oscillator at various values of the oscillator frequency $\omega_{r}$. From bottom to top, the curves correspond to: $\omega_{r}/g_{0}=30$ (black), $\omega_{r}/g_{0}=50$ (red), $\omega_{r}/g_{0}=200$ (blue),  and the analytical solution from Eq.~(\ref{avphononanal}) (purple), respectively. (b) The fidelity $f(t)$ between the states $\vert\Psi(t)\rangle$ and $\vert\psi(t)\rangle$ vs the time $t$ at the same values of $\omega_{r}$ as those in (a). (c) The fidelity $f(t_{s})$ at time $t_{s}=\pi/\delta$ vs the oscillator frequency $\omega_{r}$. Other parameters are $n_{0}=1$ and $\xi=1.5271$. We set the detuning to be $\delta=g$.}
\label{Fig3}
\end{figure}
%%%%%%%%%%%%%%%%%%%%%%%

\subsection{Fidelities of approximate solution in a closed system\label{ssec:fidelityclosesystem}}

The validity of the RWA performed in obtaining the Hamiltonian $\tilde{H}_{\mbox{\scriptsize RWA}}(t)$ can be evaluated by comparing the analytical solution in Sec.~\ref{ssec:analytical} and the numerical solution in Sec.~\ref{ssec:numerical}. First, we consider the average excitation number $\langle n_{a}(t)\rangle$ of the oscillator. For the state $\vert\Psi(t)\rangle$, we derive
\begin{eqnarray}
\langle n_{a}(t)\rangle&=&\langle\Psi(t)\vert a^{\dagger}a\vert\Psi(t)\rangle\nonumber\\
&=&\sum_{m=0}^{\infty}\left[m\left(\vert A_{m}(t)\vert^{2}+\vert B_{m}(t)\vert^{2}\right)\right].\label{avnaclonumer}
\end{eqnarray}
In Fig.~\ref{Fig3}(a), we plot the time dependence of the average excitation number $\langle n_{a}(t)\rangle$ given by Eq.~(\ref{avnaclonumer}) at selected values of the oscillator frequency $\omega_{r}$. These numerical results are compared with the analytical result given by Eq.~(\ref{avphononanal}), which is independent of $\omega_{r}$. For $\omega_{r}/g_{0}=200$, the numerical result strongly overlaps with the analytical result. We can see from the inset of this figure that the numerical results agree better with the analytical result for larger values of $\omega_{r}/g_{0}$. In these curves, the peak values of $\langle n_{a}(t)\rangle$ are located at $g_{0}t_{s}=\pi g_{0}/\delta\approx6.45$ with our parameters. This can be well explained by Eq.~(\ref{avphononanal}): at times $t=(2m+1)\pi/\delta$ for natural numbers $m$, the displacement of the oscillator reaches its maximum value of $|\alpha|_{\mbox{\scriptsize max}}=2g/\delta$.
%%%%%%%%%%%%%%%%%%%%%%%
\begin{figure}[tbp]
\center
\includegraphics[bb=28 2 438 423, width=0.47 \textwidth]{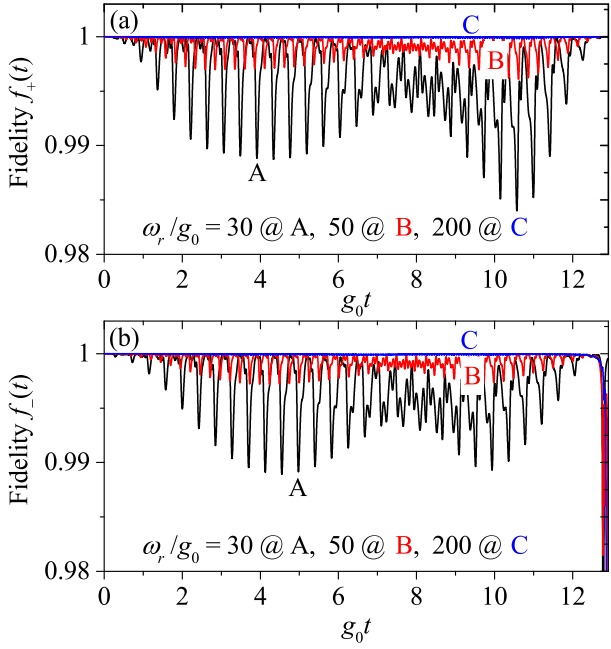}
\caption{The fidelities $f_{\pm}(t)$ between the states $\vert\Psi_{\pm}(t)\rangle$ and their corresponding target states $\vert\alpha_{\pm}(t)\rangle_{r}$ vs the time $t$. The selected values of $\omega_{r}$ are the same as those in Fig~\ref{Fig3}. Other parameters are $n_{0}=1$, $\xi=1.5271$, and $\delta=g$.}
\label{Fig4}
\end{figure}
%%%%%%%%%%%%%%%%%%%%%%%

We also examine the fidelity $f(t)=\vert\langle\Psi(t)\vert\psi(t)\rangle\vert^{2}$ between the states $\vert\Psi(t)\rangle$ and $\vert\psi(t)\rangle$. Using Eqs.~(\ref{statepsi}) and (\ref{statePsi}), this fidelity can be obtained as
\begin{eqnarray}
f(t)&=&\frac{1}{2}e^{-\vert\alpha(t)\vert^{2}}\left\vert\sum_{m=0}^{\infty}\left\{
[A_{m}^{\ast}(t)\cos[\xi\sin(\omega_{0}t)]\right.\right.\nonumber\\
&&\left.\left.-iB_{m}^{\ast}(t)\sin[\xi\sin(\omega_{0}t)]]+(-1)^{m}[B_{m}^{\ast}(t)\cos[\xi\sin(\omega_{0}t)]\right.\right.\nonumber\\
&&\left.\left.-iA_{m}^{\ast}(t)\sin[\xi\sin(\omega_{0}t)]]\right\}\frac{\alpha^{m}(t)}{\sqrt{m!}}
\right\vert ^{2}
\end{eqnarray}
in terms of the coefficients $A_{m}(t)$ and $B_{m}(t)$. In Fig.~\ref{Fig3}(b), we plot $f(t)$ with the same values of oscillator frequency as those in Fig.~\ref{Fig3}(a). The fidelity exhibits fast oscillations caused by the high-frequency oscillating phase factors $\exp(\pm i\omega_{r}t)$ and $\exp(\pm in\omega_{0}t)$ ($\omega_{r},\omega_{0}\gg g_{0}$). For a larger $\omega_{r}$, $f(t)$ oscillates faster, but with a smaller oscillation amplitude. For $\omega_{r}/g_{0}=200$, $f(t)\approx1$ with a negligible oscillating amplitude, which indicates the validity of the RWA. In Fig.~\ref{Fig3}(c) we plot the fidelity $f(t_{s})$ as a function of $\omega_{r}$ at time $t_{s}=\pi/\delta$, where $t_{s}$ corresponds to the location of the peak value in Fig.~\ref{Fig3}(a). Here $f(t_{s})$ shows a small oscillation (see inset), but with an envelope that increases gradually to reach the value $1$. This behavior agrees with the above discussions.

Now we consider the fidelities between the generated states $\vert\Psi_{\pm}(t)\rangle$ in Eq.~(\ref{stateexactclo}) and the target states $\vert\alpha_{\pm}(t)\rangle_{r}$ (the Schr\"{o}dinger-cat states) of the oscillator after a measurement on the $\sigma_{x}$ operator of the qubit is conducted. Using Eqs.~(\ref{catstates}) and~(\ref{stateexactclo}), the fidelities $f_{\pm}(t)=\vert\,_{r}\!\langle\Psi_{\pm}(t)\vert\alpha_{\pm}(t)\rangle_{r}\vert^{2}$ can be obtained as
\begin{subequations}
\begin{align}
f_{+}(t)=&\frac{2\mathcal{N}^{2}_{+}(t)e^{-|\alpha(t)|^{2}}}{p_{+}(t)}
\left\vert\sum_{l=0}^{\infty}[A_{2l}^{\ast}(t)+B_{2l}^{\ast}(t)]\frac{\alpha^{2l}(t)}{\sqrt{(2l)!}}\right\vert^{2},\\
f_{-}(t)=&\frac{2\mathcal{N}^{2}_{-}(t)e^{-|\alpha(t)|^{2}}}{p_{-}(t)}\left\vert\sum_{l=0}^{\infty}[A_{2l+1}^{\ast}(t)-B_{2l+1}^{\ast}(t)]\frac{\alpha^{2l+1}(t)}{\sqrt{(2l+1)!}}\right\vert^{2}.
\end{align}
\end{subequations}

%%%%%%%%%%%%%%%%%%%%%%%
\begin{figure}[tbp]
\center
\includegraphics[bb=1 3 433 425, width=0.47 \textwidth]{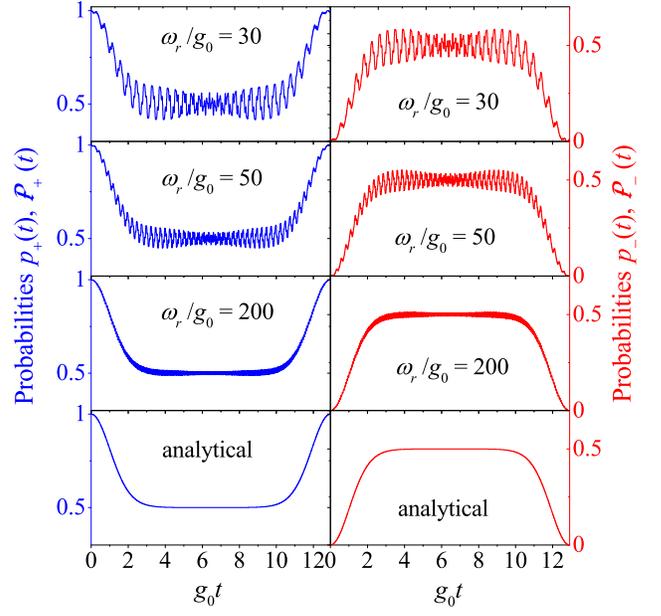}
\caption{The probabilities $p_{\pm}(t)$ of the qubit states (the upper three panels) given by Eq.~(\ref{probpmclo}) vs the time $t$ at the same values of $\omega_{r}$ as those in Fig.~\ref{Fig3}. The analytical results $\mathcal{P}_{\pm}(t)$ (the lowest panel) are given by Eq.~(\ref{Probanaly}). Other parameters are $n_{0}=1$, $\xi=1.5271$, and $\delta=g$.}
\label{Fig5}
\end{figure}
%%%%%%%%%%%%%%%%%%%%%%%
In Fig.~\ref{Fig4}, we plot the fidelities $f_{\pm}(t)$ as functions of the time. Our result shows that the fidelities increase gradually with the increase of $\omega_{r}$. One interesting effect is that around the time $t=2\pi /\delta\approx12.9$, the fidelity $f_{-}(t)$ experiences a sudden decrease. This phenomenon can be explained by analyzing the state of the oscillator at this time. When $t\rightarrow2\pi/\delta$, we have $\alpha(t)\rightarrow 0$, and then the state of the system approaches to $\vert +\rangle_{q}\vert 0\rangle_{r}$ and the target states become
\begin{equation}
\vert\alpha_{+}(2\pi/\delta)\rangle_{r}\rightarrow\vert 0\rangle_{r}, \hspace{0.5 cm} \vert\alpha_{-}(2\pi/\delta)\rangle_{r}\rightarrow\vert 1\rangle_{r}.
\end{equation}
Therefore, when $t\rightarrow2\pi/\delta$, the fidelity $f_{-}(2\pi/\delta)\rightarrow0$, if the RWA condition (\ref{RWAcond}) is well satisfied. In fact, the probability for detection of the state $|-\rangle_{q}$ at this time is also zero, as shown by Eq.~(\ref{Probanaly}).

In the upper three panels in Fig.~\ref{Fig5}, we plot the measurement probabilities $p_{\pm}(t)$ of the qubit states defined in Eq.~(\ref{probpmclo}) at selected values of $\omega_{r}$. These probabilities are compared with the analytical results $\mathcal{P}_{\pm}(t)$ given by Eq.~(\ref{Probanaly}) (the lowest panel). Figure~\ref{Fig5} shows that the probabilities from the numerical simulation of the full Hamiltonian $H(t)$ exhibit fast oscillations around the approximate values from the RWA results. With the increase of $\omega_{r}/g_{0}$, the oscillation becomes faster but its magnitude decreases gradually. The numerical results at large $\omega_{r}/g_{0}$ agree well with the RWA result. At time $t_{s}=\pi/\delta$ for a maximum oscillator displacement, the qubit will be detected in the states $|\pm\rangle_{q}$ with approximately equal probability of $1/2$.

\subsection{The Wigner function and the probability distribution of the rotated quadrature operator}

The quantum interference and coherence effects in the generated cat states can be revealed by studying the Wigner function~\cite{Buzek1995} and the probability distribution of the rotated quadrature operator~\cite{Milburnbook}.
For the harmonic oscillator described by a density matrix $\rho_{r}$, the Wigner function is defined by~\cite{Buzek1995}
\begin{eqnarray}
W(\beta)=\frac{2}{\pi}\textrm{Tr}\left[D^{\dagger}(\beta)\rho_{r} D(\beta)(-1)^{a^{\dagger}a}\right],
\end{eqnarray}
where $D(\beta)=\exp(\beta a^{\dagger}-\beta^{\ast}a)$ is a displacement operator. For the states $\vert\Psi_{\pm}(t)\rangle_{r}$ given in Eq.~(\ref{stateexactclo}), the Wigner functions can be obtained as
\begin{eqnarray}
W^{(\pm)}(\beta)&=&\frac{1}{\pi p_{\pm}(t)}\sum_{l,m,n=0}^{\infty}[A_{m}(t)\pm B_{m}(t)][A_{n}^{\ast}(t)\pm B_{n}^{\ast}(t)]\notag\\
&&\times(-1)^{l}\,_{r}\langle l\vert D^{\dagger}(\beta)\vert m\rangle_{r}\,_{r}\langle n\vert D(\beta)\vert l\rangle_{r}.
\end{eqnarray}
Here the probabilities $p_{\pm}(t)$ are given by Eq.~(\ref{probpmclo}), and the matrix elements of the displacement operator in the Fock space can be calculated by~\cite{Buek1990}
\begin{eqnarray}
_{r}\!\langle m\vert D(\beta)\vert n\rangle\!_{r}=\left\{\begin{array}{c}\sqrt{\frac{m!}{n!}}e^{-\vert\beta\vert^{2}/2}(-\beta^{\ast})^{n-m}L_{m}^{n-m}(\vert\beta\vert^{2}),\hspace{0.3 cm}n>m, \\
\sqrt{\frac{n!}{m!}}e^{-\vert\beta\vert^{2}/2}(\beta)^{m-n}L_{n}^{m-n}(\vert\beta\vert^{2}),\hspace{0.3 cm}m>n,
\end{array}\right.\label{matrixelemD}
\end{eqnarray}
where $L_{n}^{m}(x)$ are the associated Laguerre polynomials.

For the rotated quadrature operator
\begin{eqnarray}
\hat{X}(\theta)=\frac{1}{\sqrt{2}}(ae^{-i\theta}+a^{\dag}e^{i\theta}),
\end{eqnarray}
we denote the eigenstate as $|X(\theta)\rangle_{r}$: $\hat{X}(\theta)|X(\theta)\rangle_{r}=X(\theta)|X(\theta)\rangle_{r}$; then, for a density matrix $\rho_{r}$ of the oscillator, the probability distribution of the rotated quadrature operator $\hat{X}(\theta)$ is defined by~\cite{Milburnbook}
\begin{equation}
P[X(\theta)]=\,_{r}\langle X(\theta)\vert\rho_{r}\vert X(\theta)\rangle_{r}.
\end{equation}
For the states $\vert\Psi_{\pm}(t)\rangle_{r}$, we can obtain the probability distributions of the rotated quadrature
operator as
\begin{eqnarray}
P^{(\pm)}[X(\theta)]&=&_{r}\!\langle X(\theta)\vert\Psi_{\pm}(t)\rangle_{r}\,_{r}\langle\Psi_{\pm}(t)\vert X(\theta)\rangle_{r}\notag\\
&=&\frac{1}{2p_{\pm}(t)}\left\vert\sum_{m=0}^{\infty}
[A_{m}(t)\pm B_{m}(t)]\,_{r}\langle X(\theta)\vert m\rangle_{r}\right\vert^{2}.
\end{eqnarray}
Here the inner product of the number state $|m\rangle_{r}$ with the eigenstate $\vert X(\theta)\rangle_{r}$ of the rotated quadrature operator can be calculated with this relation,
\begin{equation}
_{r}\!\langle X(\theta)\vert m\rangle_{r}=\frac{H_{m}[X(\theta)]}{\sqrt{\pi^{1/2}2^{m}m!}}e^{-X^{2}(\theta)/2}e^{-i\theta m},
\end{equation}
where $H_{m}[x]$ are the Hermite polynomials.
%%%%%%%%%%%%%%%%%%%%%%%
\begin{figure}[tbp]
\center
\includegraphics[bb=1 120 586 604, width=0.47 \textwidth]{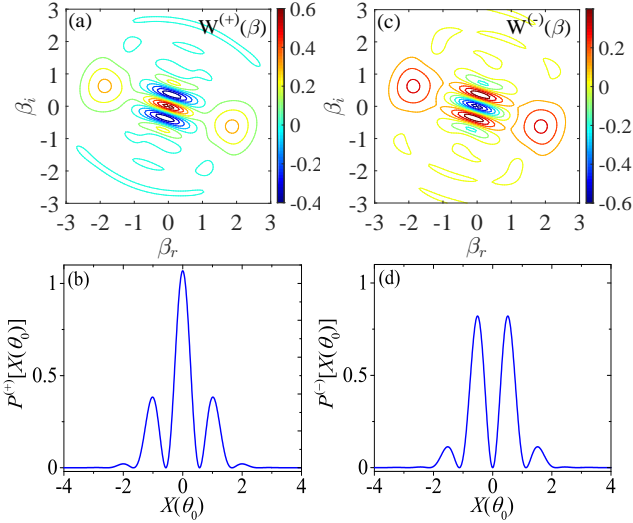}
\caption{(a),(c) The Wigner functions $W^{(\pm)}(\beta)$ and (b),(d) the probability distributions $P^{(\pm)}[X(\theta_{0})]$ of the rotated quadrature operator $\hat{X}(\theta_{0})$ for the oscillator's states $\vert\Psi_{\pm}(t_{s})\rangle_{r}$. Other parameters are $\omega_{r}/g_{0}=200$, $n_{0}=1$, $\xi=1.5271$, and $\delta=g$.}
\label{Fig6}
\end{figure}
%%%%%%%%%%%%%%%%%%%%%%%

In Fig.~\ref{Fig6}, we plot the Wigner functions $W^{(\pm)}(\beta)$ and the probability distributions $P^{(\pm)}[X(\theta_{0})]$ of the rotated quadrature operator $\hat{X}(\theta_{0})$ for the states $\vert\Psi_{\pm}(t_{s})\rangle_{r}$, where $t_{s}=\pi/\delta$ is the qubit detection time. Based on the analytical results, we know that the coherent amplitude is $\alpha(t_{s})=-1.9005+0.6228i$ [$|\alpha(t_{s})|=2$ and $\arg[\alpha(t_{s})]=2.8249$] at time $t_{s}=\pi/\delta=6.45$ and under the used parameters. In Figs.~\ref{Fig6}(a) and~\ref{Fig6}(c), we can see that the positions of the two main peaks in the Wigner functions are located at $\pm\alpha(t_{s})$ in the phase space, which represent the two coherent states $|\pm \alpha(t_{s})\rangle_{r}$. Moreover, the Wigner functions exhibit a clear oscillating pattern in the region between these two peaks (along the line that is perpendicular to the link between the two peaks). This oscillating feature is a distinct evidence of quantum interference and coherence. The quantum features can also be seen by considering the probability distribution of the rotated quadrature operator. We choose a rotated quadrature operator with the angle $\theta_{0}=\arg[\alpha(t_{s})]-\pi/2=1.2541$, which means that the quadrature direction is perpendicular to the line that links the two peaks. The interference is maximum along this direction because the two coherent states are projected onto the quadrature such that they overlap exactly. In Figs.~\ref{Fig6}(b) and~\ref{Fig6}(d), we plot $P^{(\pm)}[X(\theta_{0})]$ of the states $\vert\Psi_{\pm}(t_{s})\rangle_{r}$. Oscillating features can be seen clearly from these curves. It is interesting to point out that the pattern of the probability distribution in Figs.~\ref{Fig6}(b) and~\ref{Fig6}(d) is very similar to the curves in the cross section of the Wigner function [in Figs.~\ref{Fig6}(a) and~\ref{Fig6}(c)] at the same rotating angle. The difference is that the probability distributions are always positive; whereas the Wigner functions could be negative.

\section{The open-system case\label{sec:opensys}}
In this section, we will study how the dissipations of the system affect the fidelity and probability of the generated states. We will also discuss the influence of the dissipations on the Wigner function and the probability distribution of the rotated quadrature operator.

\subsection{Quantum master equation and solution\label{ssec:mastereq}}
%%%%%%%%%%%%%%%%%%%%%%%
\begin{figure}[tbp]
\center
\includegraphics[bb=6 1 408 362, width=0.47 \textwidth]{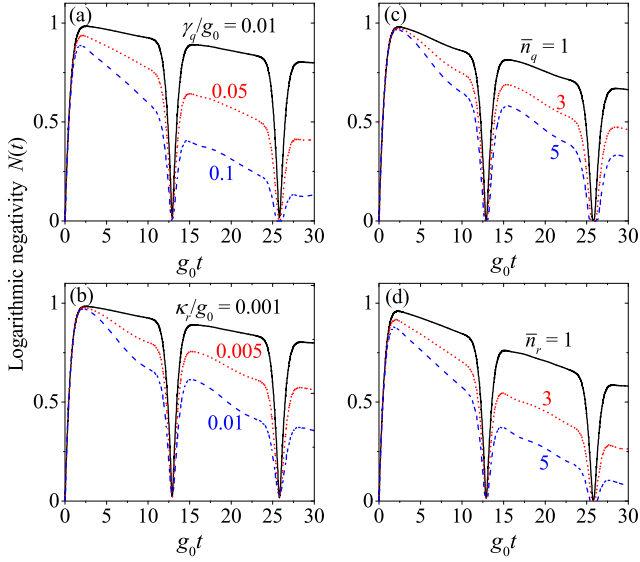}
\caption{The logarithmic negativity $N(t)$ of the density matrix $\rho(t)$ vs the time $t$ in various cases. (a) $\kappa_{r}/g_{0}=0.001$, $\bar{n}_{r}=0$, $\bar{n}_{q}=0$, and $\gamma_{q}/g_{0}=0.01$, $0.05$, and $0.1$; (b) $\gamma_{q}/g_{0}=0.01$, $\bar{n}_{q}=0$, $\bar{n}_{r}=0$, and $\kappa_{r}/g_{0}=0.001$, $0.005$, and $0.01$; (c) $\kappa_{r}/g_{0}=0.001$, $\bar{n}_{r}=0$, $\gamma_{q}/g_{0}=0.01$, and $n_{q}=1$, $3$, and $5$; (d) $\gamma_{q}/g_{0}=0.01$, $\bar{n}_{q}=0$, $\kappa_{r}/g_{0}=0.001$, and $\bar{n}_{r}=1$, $3$, and $5$. Other parameters are $\omega_{r}/g_{0}=200$, $n_{0}=1$, $\xi=1.5271$, and $\delta=g$.}
\label{Fig7}
\end{figure}
%%%%%%%%%%%%%%%%%%%%%%%
Under environmental fluctuations, the evolution of the system is governed by the quantum master equation,
\begin{eqnarray}
\dot{\rho}&=&i[\rho,H(t)]+\gamma_{q}\bar{n}_{q}\mathcal{D}[\sigma_{+}]\rho+\gamma_{q}(\bar{n}_{q}+1)\mathcal{D}[\sigma_{-}]\rho\nonumber\\
&&+\kappa_{r}\bar{n}_{r}\mathcal{D}[a^{\dagger}]\rho+\kappa_{r}(\bar{n}_{r}+1)\mathcal{D}[a]\rho,\label{densmatgen}
\end{eqnarray}
where $\mathcal{D}[o]\rho=o\rho o^{\dagger}-(o^{\dagger}o\rho+\rho o^{\dagger}o)/2$ is the standard Lindblad superoperator that describes the dampings of the qubit and the oscillator. The parameters $\gamma_{q}$ ($\kappa_{r}$) and $\bar{n}_{q}$ ($\bar{n}_{r}$) are the damping rate and the thermal excitation number for the qubit (oscillator) bath, respectively. To solve this master equation, we expand the state of the oscillator in the Fock space. Then the density matrix of the total system can be expressed as
\begin{eqnarray}
\rho(t)&=&\sum_{m,n=0}^{\infty}[\rho_{1,m,1,n}(t)\vert 1\rangle_{q}\vert m\rangle_{r}\;_{q}\langle 1\vert_{r}\langle n\vert\nonumber\\
&&+\rho_{1,m,0,n}(t)\vert 1\rangle_{q}\vert m\rangle_{r}\;_{q}\langle 0\vert_{r}\langle n\vert\nonumber\\
&&+\rho_{0,m,1,n}(t)\vert 0\rangle_{q}\vert m\rangle_{r}\;_{q}\langle 1\vert_{r}\langle n\vert\nonumber\\
&&+\rho_{0,m,0,n}(t)\vert 0\rangle_{q}\vert m\rangle_{r}\;_{q}\langle 0\vert_{r}\langle n\vert].\label{denmatrrho}
\end{eqnarray}
For an initial state $\vert +\rangle_{q}\vert 0\rangle_{r}$, the nonzero density matrix elements are $\rho_{0,0,0,0}(0)=\rho_{0,0,1,0}(0)=\rho_{1,0,0,0}(0)=\rho_{1,0,1,0}(0)=1/2$. By numerically solving the master equation (\ref{densmatgen}) under the initial condition, the time evolution of the density matrix $\rho(t)$ can be obtained.

\subsection{Entanglement between the qubit and the oscillator}

In the open-system case, the entanglement of the density matrix $\rho(t)$ can be quantified by calculating the logarithmic negativity.
In terms of Eqs.~(\ref{lognegtivity}), (\ref{densmatgen}), and~(\ref{denmatrrho}), the logarithmic negativity of the state $\rho(t)$ can be obtained numerically.
In Fig.~\ref{Fig7}, we plot the logarithmic negativity $N(t)$ as a function of time $t$ when the dissipation parameters of the system
take various values. Here, Figs.~\ref{Fig7}(a) and~\ref{Fig7}(c) are plotted for various values of $\gamma_{q}$ and $\bar{n}_{q}$, while Figs.~\ref{Fig7}(b) and~\ref{Fig7}(d) show the curves corresponding to different values of $\kappa_{r}$ and $\bar{n}_{r}$.
Similar to the pure-state case in Fig.~\ref{Fig2}, at the decoupling times $t_{m}=2m\pi/\delta$ for natural numbers $m$, the qubit and the oscillator decouple and the logarithmic negativity becomes zero. In the middle durations between neighboring decoupling times, the logarithmic negativity decreases with time. When the values around the decoupling times are ignored, the logarithmic negativity in these middle durations exhibits a tendency to smoothly decrease. For larger values of decay rates and thermal excitation numbers, the logarithmic negativity decays faster.

\subsection{Fidelity and probability of the cat states}

%%%%%%%%%%%%%%%%%%%%%%%
\begin{figure}[tbp]
\center
\includegraphics[bb=3 1 526 455, width=0.47 \textwidth]{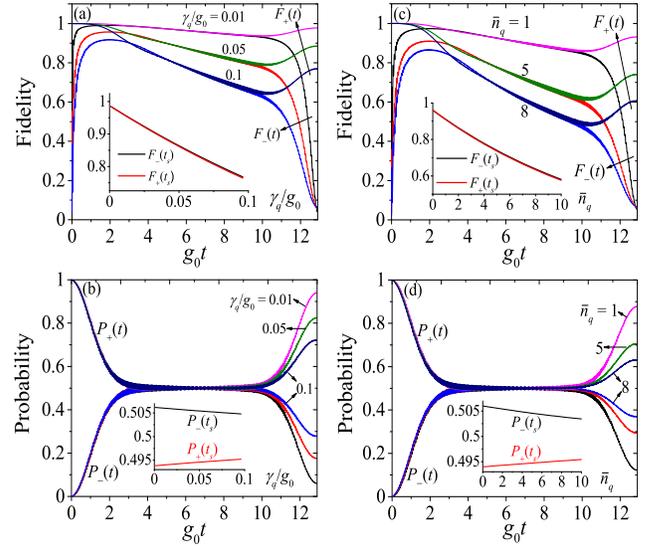}
\caption{The fidelities $F_{\pm}(t)$ and the probabilities $P_{\pm}(t)$ as functions of time $t$ at (a),(b) $\bar{n}_{q}=0$ and $\gamma_{q}/g_{0}=0.01$, $0.05$, and $0.1$; (c),(d) $\gamma_{q}/g_{0}=0.01$ and $\bar{n}_{q}=1$, $5$, and $8$. Other parameters are $\omega_{r}/g_{0}=200$, $n_{0}=1$, $\xi=1.5271$, $\delta=g$, $\kappa_{r}/g_{0}=0.001$, and $\bar{n}_{r}=0$. The insets are the fidelities $F_{\pm}(t_{s})$ and the probabilities $P_{\pm}(t_{s})$ at $t_{s}=\pi/\delta$ vs $\gamma_{q}$ and $\bar{n}_{q}$.}
\label{Fig8}
\end{figure}
%%%%%%%%%%%%%%%%%%%%%%%
As explained in Sec.~\ref{sec:stategeneration}, in order to create quantum superpositions in the oscillator, we need to perform a projective measurement on the qubit. For a density matrix $\rho(t)$, when the qubit is detected in states $\vert\pm\rangle_{q}$, the reduced density matrices of the oscillator are
\begin{equation}
\rho_{r}^{(\pm)}(t)=\frac{1}{2P_{\pm}(t)}\sum_{m,n=0}^{\infty}\Lambda^{(\pm)}_{m,n}(t)\vert m\rangle_{r}\!\,_{r}\!\langle n\vert,\label{mixedstate}
\end{equation}
where we introduced the variables
\begin{eqnarray}
\Lambda^{(\pm)}_{m,n}(t)&=&\rho_{1,m,1,n}(t)+\rho_{0,m,0,n}(t)\nonumber\\
&&\pm[\rho_{1,m,0,n}(t)+\rho_{0,m,1,n}(t)],
\end{eqnarray}
and the probabilities for detecting the qubit states $\vert\pm\rangle_{q}$,
\begin{equation}
P_{\pm}(t)=\frac{1}{2}\sum_{m=0}^{\infty}\Lambda^{(\pm)}_{m,m}(t).\label{Prodpmopen}
\end{equation}
We can evaluate the efficiency of the state generation by calculating the fidelities $F_{\pm}(t)=\,_{r}\langle\alpha_{\pm}(t)\vert\rho_{r}^{(\pm)}(t)\vert\alpha_{\pm}(t)\rangle_{r}$ between the generated states $\rho_{r}^{(\pm)}(t)$ and the target states $\vert \alpha_{\pm}(t)\rangle_{r}$. The fidelities have the form
\begin{subequations}
\label{Fidelityopen}
\begin{align}
F_{+}(t)=&\frac{2\mathcal{N}^{2}_{+}(t)e^{-\vert\alpha(t)\vert^{2}}}{P_{+}(t)}\sum_{l,l^{\prime}=0}^{\infty}\Lambda^{(+)}_{2l^{\prime},2l}(t)
\frac{[\alpha^{\ast}(t)]^{2l^{\prime}}\alpha^{2l}(t)}{\sqrt{(2l^{\prime})!(2l)!}},\\
F_{-}(t)=&\frac{2\mathcal{N}^{2}_{-}(t)e^{-\vert\alpha(t)\vert^{2}}}{P_{-}(t)}\sum_{l,l^{\prime}=0}^{\infty}\Lambda^{(-)}_{2l^{\prime}+1,2l+1}(t)\frac{[\alpha ^{\ast}(t)]^{(2l^{\prime}+1)}\alpha^{2l+1}(t)}{\sqrt{(2l^{\prime }+1)!(2l+1)!}}.
\end{align}
\end{subequations}
%%%%%%%%%%%%%%%%%%%%%%%
\begin{figure}[tbp]
\center
\includegraphics[bb=3 1 529 455, width=0.47 \textwidth]{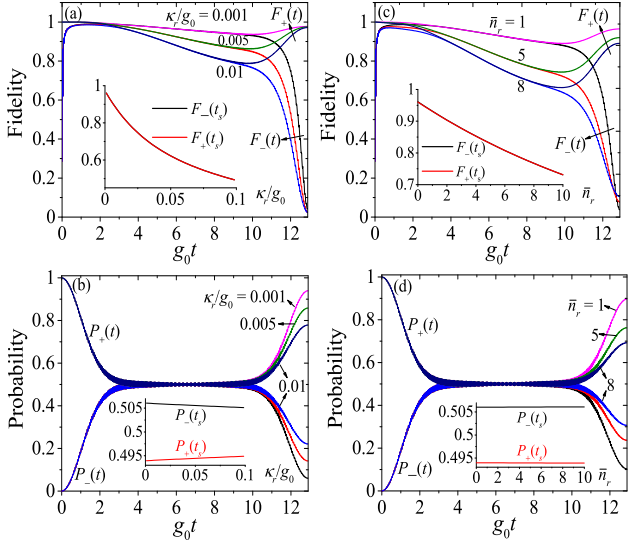}
\caption{The fidelities $F_{\pm}(t)$ and the probabilities $P_{\pm}(t)$ as functions of time $t$ at (a),(b) $\bar{n}_{r}=0$ and $\kappa_{r}/g_{0}=0.001$, $0.005$, and $0.01$; (c),(d) $\kappa_{r}/g_{0}=0.001$ and $\bar{n}_{r}=1$, $5$, and $8$. Other parameters are $\omega_{r}/g_{0}=200$, $n_{0}=1$, $\xi=1.5271$, $\delta=g$, $\gamma_{q}/g_{0}=0.01$, and $\bar{n}_{q}=0$. The insets are the fidelities $F_{\pm}(t_{s})$ and the probabilities $P_{\pm}(t_{s})$ at $t_{s}=\pi/\delta$ versus $\kappa_{r}$ and $\bar{n}_{r}$.}
\label{Fig9}
\end{figure}
%%%%%%%%%%%%%%%%%%%%%%%

In Fig.~\ref{Fig8}, we display the time dependence of the fidelities $F_{\pm}(t)$ and the probabilities $P_{\pm}(t)$ at selected values of qubit decay rate $\gamma_{q}$ and thermal excitation number $\bar{n}_{q}$. In the intermediate duration of $g_{0}t\approx 3$ - $10$ [corresponding to $|\alpha(t)|>1.3$ and $\exp(-2|\alpha(t)|^{2})<0.034$], the fidelities $F_{+}(t)$ and $F_{-}(t)$ have approximately equal values, and the probabilities $P_{+}(t)\approx P_{-}(t)\approx1/2$. At a given time in this duration, $F_{\pm}(t)$ decrease with the increase of $\gamma_{q}$ and $\bar{n}_{q}$. This feature can be seen more clearly at time $t_{s}=\pi/\delta$ when the displacement reaches its peak values. As shown in the insets of Figs.~\ref{Fig8}(a) and~\ref{Fig8}(c), $F_{\pm}(t_{s})$ decrease rapidly with the increase of $\gamma_{q}$ and $\bar{n}_{q}$. On the contrary, the probability $P_{+}(t_{s})$ [$P_{-}(t_{s})$] at the peak only increases (decreases) very slowly with the increase of $\gamma_{q}$ and $\bar{n}_{q}$, under the normalization $P_{+}(t_{s})+P_{-}(t_{s})=1$ [the insets in Figs.~\ref{Fig8}(b) and~\ref{Fig8}(d)]. Around the times $g_{0}t=0$ and $12.91$, $F_{+}(t)$ and $F_{-}(t)$ have different values. Here, $F_{+}(t\rightarrow0)=1$ because $\vert\alpha_{+}(t\rightarrow0)\rangle_{r}=\vert 0\rangle_{r}$ is exactly the initial state. The value $F_{-}(t\rightarrow0)=0$ because $\vert\alpha_{-}(t\rightarrow0)\rangle_{r}=|1\rangle_{r}$. The corresponding probability $P_{-}(0)=0$ in the ideal case. Approaching the time $g_{0}t=12.91$, the fidelity $F_{+}(t)$ [$F_{-}(t)$] has a tendency of increasing (decreasing). The probabilities $P_{\pm}(t)$ have small deviation from the analytical results in Fig.~\ref{Fig5}, and the deviation increases with $\gamma_{q}$ and $\bar{n}_{q}$.

In Fig.~\ref{Fig9}, we plot the fidelities $F_{\pm}(t)$ and the probabilities $P_{\pm}(t)$ at various values of the oscillator decay rate $\kappa_{r}$ and the thermal excitation number $\bar{n}_{r}$, as well as the fidelities $F_{\pm}(t_{s})$ and the probabilities $P_{\pm}(t_{s})$ at the peak position $t_{s}$. We can see similar behavior as that in Fig.~\ref{Fig8}. During the time period with $|\alpha(t)|>1$, the fidelities $F_{+}(t)$ and $F_{-}(t)$ have approximately equal values, and they decrease with the increase of $\kappa_{r}$ and $\bar{n}_{r}$. The probabilities $P_{+}(t)$ and $P_{-}(t)$ are approximately equal to $1/2$. The probabilities and fidelities at time $t_{s}$ are also similar to those in Fig.~\ref{Fig8}.

\subsection{The Wigner function and the probability distribution of the rotated quadrature operators\label{ssec:distribution}}
%%%%%%%%%%%%%%%%%%%%%%%
\begin{figure}[tbp]
\center
\includegraphics[bb=26 8 440 550, width=0.47 \textwidth]{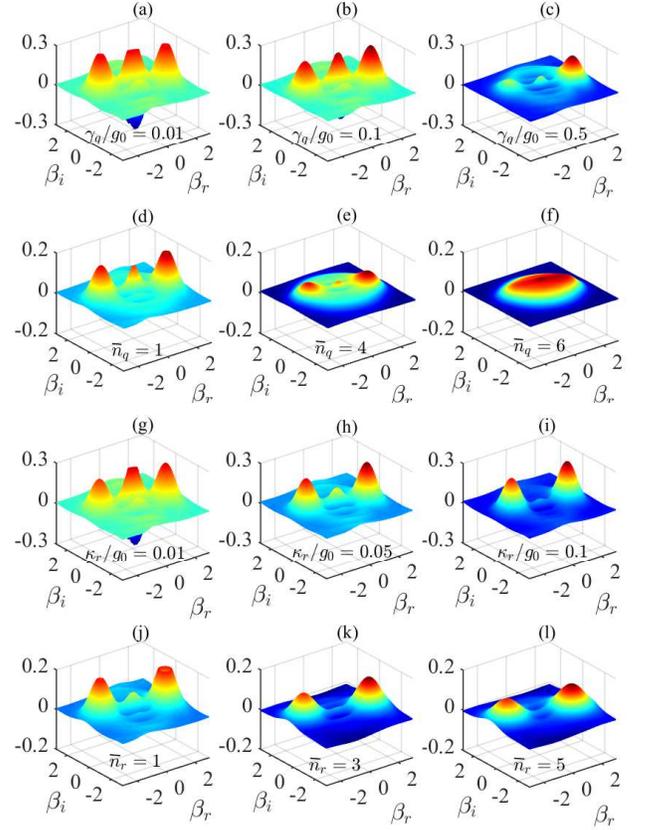}
\caption{The Wigner function $W^{(+)}(\beta)$ for the state $\rho_{r}^{(+)}(t_{s})$ in various cases. (a)-(c) $\kappa_{r}/g_{0}=0.02$, $\bar{n}_{r}=0$, $\bar{n}_{q}=0$, and $\gamma_{q}/g_{0}=0.01$, $0.1$, $0.5$; (d)-(f) $\kappa_{r}/g_{0}=0.02$, $\bar{n}_{r}=0$, $\gamma_{q}/g_{0}=0.1$, and $\bar{n}_{q}=1$, $4$, $6$; (g)-(i) $\gamma_{q}/g_{0}=0.1$, $\bar{n}_{q}=0$, $\bar{n}_{r}=0$, and $\kappa_{r}/g_{0}=0.01$, $0.05$, $0.1$; (j)-(l) $\gamma_{q}/g_{0}=0.1$, $\bar{n}_{q}=0$, $\kappa_{r}/g_{0}=0.02$, and $\bar{n}_{r}=1$, $3$, $5$. Other parameters are $\omega_{r}/g_{0}=200$, $n_{0}=1$, $\xi=1.5271$, and $\delta=g$.}
\label{Fig10}
\end{figure}
%%%%%%%%%%%%%%%%%%%%%%%
In the open-system case, the dissipation of the system will attenuate the macroscopic quantum coherence in the generated cat states. For the density matrices~(\ref{mixedstate}) of the oscillator, the Wigner functions can be obtained as
\begin{eqnarray}
W^{(\pm)}(\beta)&=&\frac{1}{\pi P_{\pm}(t)}\sum_{l,m,n=0}^{\infty}(-1)^{l}\Lambda^{(\pm)}_{m,n}(t)\,_{r}\nonumber\\
&&\times\!\langle l\vert D^{\dag}(\beta)\vert m\rangle_{r}\,_{r}\langle n\vert D(\beta)\vert l\rangle_{r},
\end{eqnarray}
where the matrix elements of the displacement operator are calculated with Eq.~(\ref{matrixelemD}).

In Fig.~\ref{Fig10} we plot the Wigner function $W^{(+)}(\beta)$ of the density
matrix $\rho_{r}^{(+)}(t_{s})$ when the decay rates and the thermal excitation numbers take various values.
Here we only show the Wigner function $W^{(+)}(\beta)$ for concision because $W^{(-)}(\beta)$ has a similar parameter dependence.
We can see from Fig.~\ref{Fig10} that, with the increase of the decay rate $\gamma_{q}$ ($\kappa_{r}$)
and the thermal excitation number $\bar{n}_{q}$ ($\bar{n}_{r}$) of the qubit (oscillator),
the interference pattern (the region between these two peaks) in the Wigner function attenuates gradually, which means that the dissipations of the qubit and the oscillator will hurt the macroscopic quantum coherence.

\begin{figure}[tbp]
\center
\includegraphics[bb=1 1 423 347, width=0.47 \textwidth]{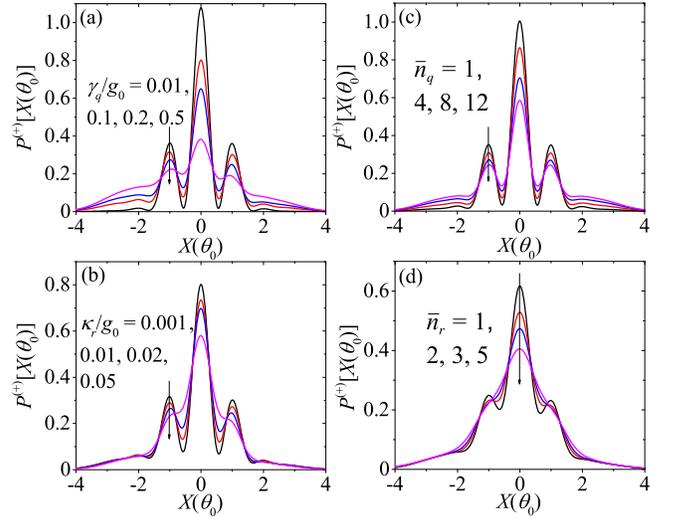}
\caption{The probability distribution $P^{(+)}[X(\theta_{0})]$ of the rotated quadrature operator $\hat{X}(\theta_{0})$ for the state $\rho_{r}^{(+)}(t_{s})$ in various cases: (a) $\kappa_{r}/g_{0}=0.001$, $\bar{n}_{r}=0$, $\bar{n}_{q}=0$, and $\gamma_{q}/g_{0}=0.01$, $0.1$, $0.2$, $0.5$; (b) $\gamma_{q}/g_{0}=0.1$, $\bar{n}_{q}=0$, $\bar{n}_{r}=0$, and $\kappa_{r}/g_{0}=0.001$, $0.01$, $0.02$, $0.05$;
(c) $\kappa_{r}/g_{0}=0.001$, $\bar{n}_{r}=0$, $\gamma_{q}/g_{0}=0.01$, and $\bar{n}_{q}=1$, $4$, $8$, $12$; (d) $\gamma_{q}/g_{0}=0.1$, $\bar{n}_{q}=0$, $\kappa_{r}/g_{0}=0.01$, and $\bar{n}_{r}=1$, $2$, $3$, $5$. Other parameters are $\omega_{r}/g_{0}=200$, $n_{0}=1$, $\xi=1.5271$, and $\delta=g$.}
\label{Fig11}
\end{figure}

We also investigate how the dissipations of the system affect the probability distributions of the rotated quadrature operator. For states~(\ref{mixedstate}), the probability distributions of the rotated quadrature operator $\hat{X}(\theta)$ can be obtained as
\begin{eqnarray}
P^{(\pm)}[X(\theta)]&=&\frac{e^{-X^{2}(\theta)}}{2P_{\pm}(t)}\sum_{m,n=0}^{\infty}\frac{\Lambda^{(\pm)}_{m,n}(t)}{\sqrt{\pi 2^{m+n}m!n!}}\nonumber\\
&&\times H_{m}[X(\theta)]H_{n}[X(\theta)] e^{i\theta(n-m)}.
\label{PDcatstate}
\end{eqnarray}
In Fig.~\ref{Fig11}, we plot the probability distribution $P^{(+)}[X(\theta_{0})]$ for the state $\rho_{r}^{(+)}(t_{s})$ at selected values of the decay rate $\gamma_{q}$ ($\kappa_{r}$) and the thermal excitation number $\bar{n}_{q}$ ($\bar{n}_{r}$) of the qubit (oscillator). It can be seen that with the increase of the four parameters, the oscillation amplitude of the probability distribution decreases gradually. This means that the dissipations of the qubit and the oscillator attenuate the quantum interference and coherence in these cat states. We also studied the influence of the system's dissipations on the probability distribution $P^{(-)}[X(\theta_{0})]$ for the state $\rho_{r}^{(-)}(t_{s})$, and a similar parameter dependence can be seen in this case.
%%%%%%%%%%%%%%%%%%%%%%%
%%%%%%%%%%%%%%%%%%%%%%%

\section{Other types of qubit-oscillator coupling\label{sec:othercouplings}}

Our approach can be applied to various other types of qubit-oscillator coupling. One example is a qubit-oscillator system with the Hamiltonian,
\begin{eqnarray}
H_{1}(t)&=&\xi\omega_{0}\cos(\omega_{0}t)\sigma_{z}+\omega_{r}a^{\dagger}a+g_{0}\sigma_{x}\left(a+a^{\dagger}\right).\label{Hamilt1rot}
\end{eqnarray}
In this model, the displacement of the oscillator can be amplified by the driving on the qubit $\xi\omega_{0}\cos(\omega_{0}t)\sigma_{z}$, which can be easily proved with the same procedure as that in Sec.~\ref{sec:model} by replacing $\sigma_{z}$ in the transformation [Eq.~(\ref{transfV})] with $\sigma_{x}$ and assuming $\omega_{q}=0$.

Another example is a Jaynes-Cummings type of coupling described by the Hamiltonian,
\begin{eqnarray}
H_{2}(t)&=&\frac{\omega_{q}}{2}\sigma_{z}+\xi \omega_{0}\cos(\omega_{0}t)\sigma_{x}+\omega_{r}a^{\dagger}a\nonumber\\
&&+g_{0}\left(\sigma_{+}a+\sigma_{-}a^{\dagger}\right).
\end{eqnarray}
Here the qubit is driven by a $\sigma_x$ field with frequency $\omega_0$. We can use the transformation $V(t)$ given in Eq.~(\ref{transfV}). In the rotating frame defined by $V(t)$, the Hamiltonian becomes
\begin{eqnarray}
\tilde{H}_{2}(t)&=&V^{\dag}(t)H_{2}(t)V(t)-iV^{\dag}(t)\dot{V}(t)\nonumber\\
&=&\frac{\omega_{q}}{2}\left\{\cos[2\xi\sin(\omega_{0}t)]\sigma_{z}+\sin[2\xi\sin(\omega_{0}t)]\sigma_{y}\right\}\nonumber\\
&&+i\frac{g_{0}}{2}\cos[2\xi\sin(\omega_{0}t)]\sigma_{y}\left(ae^{-i\omega_{r}t}-a^{\dagger}e^{i\omega_{r}t}\right)\nonumber\\
&&-i\frac{g_{0}}{2}\sin[2\xi\sin(\omega_{0}t)]\sigma_{z}\left(ae^{-i\omega_{r}t}-a^{\dagger}e^{i\omega_{r}t}\right)\nonumber\\
&&+\frac{g_{0}}{2}\sigma_{x}\left(ae^{-i\omega_{r}t} +a^{\dagger}e^{i\omega_{r}t}\right).
\end{eqnarray}
In terms of the Jacobi-Anger expansions in Eq.~(\ref{JAexpansion}), we can decompose $\tilde{H}_{2}(t)$ into different frequency components. Under the condition,
\begin{equation}
\omega_{q}/2\ll\omega_{0},\hspace{0.8 cm}\delta' \leq g', \hspace{0.8 cm}g_{0}\ll \omega_{0},\;\omega_{r},\label{condJCcase}
\end{equation}
where $\delta'=\omega_{r}-(2n_{0}-1)\omega_{0}$ and $g'=g_{0}J_{2n_{0}-1}(2\xi)/2$, we obtain an approximate Hamiltonian
\begin{equation}
\tilde{H}_{\mbox{\scriptsize RWA}}^{(2)}(t)\approx\frac{\omega_{q}}{2}J_{0}(2\xi)\sigma_{z}-g'\sigma_{z}\left(ae^{-i\delta' t}+a^{\dagger}e^{i\delta' t}\right)
\end{equation}
by applying the RWA. In this case, the maximum displacement of the oscillator is $2g'/\delta'$, which is significantly enhanced by choosing a small $\delta'$. Based on this Hamiltonian, one can prove the generation of macroscopic Schr\"{o}dinger's cat states in the oscillator using the same procedure as that in Sec.~\ref{sec:stategeneration}.

\section{Discussions \label{sec:discussion}}

To implement the above scheme, one needs appropriate procedures for initial-state preparation, qubit-state detection, and quantum tomography of the oscillator  state. The superposition state of the qubit can be easily prepared by applying pulses to rotate the qubit state.  By driving the qubit with a $\pi/2$ pulse, the qubit can evolve from the states $|0\rangle_{q}$ and $|1\rangle_{q}$ to the states $|\pm\rangle_{q}$. The ground state of the oscillator can be prepared with ground-state cooling techniques that have been realized in several experiments~\cite{Teufel2011,Chan2011}. Various approaches to measuring the qubit have been developed and have been demonstrated in experiments. The superposed states of the oscillator can be measured with quantum-state tomography schemes~\cite{Raymer2009} for electrical~\cite{Lutterbach1997} or mechanical~\cite{Rabl2004} resonators.

One potential experimental system to demonstrate the proposed scheme is a superconducting qubit coupled to a nanomechanical resonator~\cite{LaHaye2009}. For example, we can have a transmon qubit coupled to a nanomechanical resonator capacitively. This system is described by Hamiltonian~(\ref{Hamilt1}). Realistic parameters of this system could be: $\omega_{r}=2\pi\times 58$ MHz, $g_{0}\approx 2\pi\times0.3$ - $2.3$ MHz, and $\kappa_{r}\approx 2\pi\times0.967$ - $1.934$ kHz. Here the ratio $\omega_{r}/g_{0}$ is in the range of $20$ - $200$. By varying the gate voltage, we can have $\omega_{q}=0$. The values of $\delta$, $\xi$, and $n_{0}$ can be chosen to satisfy the RWA condition (\ref{RWAcond}) by using a well-designed magnetic flux threading through the superconducting quantum interference device loop of the qubit. To observe macroscopic quantum coherence, the influence of dissipation should be negligible during the whole process of the state generation. When we choose $g_{0}\approx 2\pi\times2.3$ MHz, $\delta=g$, and $\xi=1.5271$, the state-generation time is $t_{s}=\pi/\delta\approx0.45$ $\mu$s. In the nanomechanical system, the decay rate of the resonator is much smaller than the coupling strength with $\kappa_{r}/g_{0}\approx10^{-3}$. With thermal phonon number on the order of $10$,  the mechanical dissipation will not strongly affect the quantum coherence. Meanwhile, the life time of superconducting qubits in current technology can reach $100\,\mu\textrm{s}$, far exceeding the duration of the state-generation scheme~\cite{Rigetti2012}. These parameters show that it is promising to observe macroscopic quantum coherence in the proposed state-generation scheme.

\section{Conclusions\label{sec:conclusion}}
To conclude, we proposed a scheme to generate macroscopic Schr\"{o}dinger-cat states in a generic coupled qubit-oscillator system. The scheme is realized by introducing a monochromatic driving on the qubit that is coupled to the oscillator by a conditional displacement interaction. Under appropriate conditions, the driving on the qubit induces an effective resonant or near-resonant force acting on the oscillator, which can amplify the displacement of the oscillator to exceed the amplitude of quantum fluctuations. We studied the state preparation process in detail in both the closed- and open-system cases. We also studied the quantum interference and coherence in the generated states by calculating the Wigner function and the probability distribution of the rotated quadrature operator. Our results show that the proposed method could be a realistic scheme to generate strong quantum superposition in macroscopic systems.

\begin{acknowledgments}
J.Q.L and L.T. are supported by the National Science Foundation under Award No. NSF-DMR-0956064. J.F.H. is supported by the National Natural Science Foundation of China under Grants No. 11447102 and No. 11505055.
\end{acknowledgments}

\appendix*
\section{Derivation of Eq.~(\ref{unitaryoptana})}
In this appendix, we present a detailed derivation of the unitary evolution operator $U(t)$ given in Eq.~(\ref{unitaryoptana}).
For the Hamiltonian $\tilde{H}_{\textrm{RWA}}(t)=g\sigma_{z}(ae^{-i\delta t}+a^{\dagger}e^{i\delta t})$,
the unitary evolution operator $U(t)$ is determined by the equation
$i\partial U(t)/\partial t=\tilde{H}_{\textrm{RWA}}(t)U(t)$
subject to the initial condition $U(0)=I$, where $I$ is the identity matrix in the Hilbert space of the system. Formally, we can express the operator $U(t)$ as
\begin{equation}
U(t)=\mathcal{T}\exp\left\{\int_{0}^{t}\left[-i\tilde{H}_{\textrm{RWA}}(\tau)\right]d\tau\right\},
\end{equation}
where $\mathcal{T}$ is the time-ordering operator.
According to the Magnus theory, the operator $U(t)$ can be expressed as
\begin{eqnarray}
U(t)=\exp[\Lambda(t)],\hspace{1 cm}  \Lambda(t)=\sum\limits_{k=1}^{\infty}\Lambda_{k}(t).
\end{eqnarray}
Here the variables $\Lambda_{k}(t)$ are defined by
\begin{eqnarray}
\Lambda_{1}(t)&=&\int_{0}^{t}[-i\tilde{H}_{\textrm{RWA}}(t_{1})]dt_{1},\nonumber\\
\Lambda_{2}(t)&=&\frac{1}{2}\int_{0}^{t}dt_{1}\int_{0}^{t_{1}}dt_{2}[-i\tilde{H}_{\textrm{RWA}}(t_{1}),-i\tilde{H}_{\textrm{RWA}}(t_{2})],\nonumber\\
\Lambda_{3}(t)&=&\frac{1}{6}\int_{0}^{t}dt_{1}\int_{0}^{t_{1}}dt_{2}\int_{0}^{t_{2}}dt_{3}\nonumber\\
&&\times\left([-i\tilde{H}_{\textrm{RWA}}(t_{1}),[-i\tilde{H}_{\textrm{RWA}}(t_{2}),-i\tilde{H}_{\textrm{RWA}}(t_{3})]]\right.\nonumber\\
&&\left.+[-i\tilde{H}_{\textrm{RWA}}(t_{3}),[-i\tilde{H}_{\textrm{RWA}}(t_{2}),-i\tilde{H}_{\textrm{RWA}}(t_{1})]]\right),\nonumber\\
\Lambda_{k>3}(t)&=&\dots.
\end{eqnarray}
Here $[A, B]=AB-BA$ is the matrix commutator of $A$ and $B$, and the higher-order terms consist of the integral of the commutators of Hamiltonians at different times. Since the commutator of two Hamiltonians at different times is a c-number, i.e.,
\begin{eqnarray}
&&\left[-i\tilde{H}_{\textrm{RWA}}( t'),-i\tilde{H}_{\textrm{RWA}}(t'')\right] =2ig^{2}\sin \left[ \delta \left( t'-t''\right) \right],
\end{eqnarray}
the third- and higher-order terms in the Magnus expansion vanish, $\Lambda_{k>2}(t)=0$. Using the Hamiltonian $\tilde{H}_{\textrm{RWA}}(t)$, we obtain
\begin{eqnarray}
\Lambda_{1}\left( t\right) &=&\sigma _{z}\left[\frac{g}{\delta }\left(1-e^{i\delta t}\right)
a^{\dagger}-\frac{g}{\delta }\left(1-e^{-i\delta t}\right)a\right],\nonumber\\
\Lambda_{2}\left( t\right) &=&i\frac{g^{2}}{\delta^{2}}\left[\delta t-\sin\left(\delta t\right)\right].
\end{eqnarray}
Then the operator $U(t)$ can be expressed by Eq.~(\ref{unitaryoptana}).

\end{document}